\documentclass[submission,copyright,creativecommons]{eptcs}
\providecommand{\event}{MSFP 2022} 

\usepackage{comment}
\usepackage{underscore}
\usepackage{amsmath}
\usepackage{amssymb}
\usepackage{amsthm}

\newcommand{\Nat}{\mathbb{N}}
\newcommand{\Int}{\mathbb{Z}}
\newcommand{\IntInf}{\mathbb{Z}_{\infty}}
\newcommand{\Rat}{\mathbb{Q}}
\newcommand{\Real}{\mathbb{R}}
\newcommand{\Complex}{\mathbb{C}}

\newcommand{\zerok}{\mathbf{0}}
\newcommand{\free}[2]{\mathbf{F}_{#1}[#2]}
\newcommand{\freek}[1]{\free{\K}{#1}}
\newcommand{\cfree}[2]{\mathbf{F}^*_{#1}[#2]}
\newcommand{\cfreek}[1]{\cfree{\K}{#1}}
\newcommand{\fmap}{\Rightarrow}
\newcommand{\cmap}{\Rightarrow^*}

\newcommand{\car}[1]{| #1 |}
\newcommand{\ext}[1]{\,\widehat{{#1}}\,}

\newcommand{\ssum}{\textstyle\sum}
\newcommand{\weight}{\#}

\newcommand{\inl}{\mathbf{left}}
\newcommand{\inr}{\mathbf{right}}

\newcommand{\id}[1]{\mathrm{id}_{#1}}
\newcommand{\iso}{\cong}
\newcommand{\isodef}{\leftrightarrow}
\newcommand{\fmapemptyiso}{\mathbf{cp}_0}
\newcommand{\fmapunitiso}{\mathbf{cp}_1}
\newcommand{\fmapsumiso}{\mathbf{cp}_+}
\newcommand{\fmapprodiso}{\mathbf{cp}_{\times}}

\newcommand{\return}[1]{\langle {#1} \rangle}

\newcommand{\Null}{\mathbf{null}}
\newcommand{\comp}{\circ}

\newcommand{\Str}{\mathbf{Str}}
\newcommand{\inj}[1]{\return{#1}}
\newcommand{\casemap}[1]{\mathbf{case}(#1)}

\newcommand{\K}{K}

\newcommand{\FinField}[1]{\mathbb{F}_{#1}}

\newcommand{\s}[1]{\{#1\}}

\newtheorem{definition}{Definition}

\begin{document}

\title{The Programming of Algebra}         

\author{Fritz Henglein
\institute{DIKU, Department of Computer Science, \\ University of Copenhagen}
\and 
Robin Kaarsgaard
\institute{School of Informatics, \\ University of Edinburgh}
\and 
Mikkel Kragh Mathiesen
\institute{DIKU, Department of Computer Science, \\ University of Copenhagen}}

\def\authorrunning{Henglein, Kaarsgaard, \& Mathiesen}
\def\titlerunning{The Programming of Algebra}

\maketitle

\begin{abstract}

We present module theory and linear maps as a powerful generalised and computationally efficient framework for the relational data model, which underpins today's relational database systems.
Based on universal constructions of modules we obtain compact and computationally efficient data structures for data collections 
corresponding to union and deletion, repeated union, Cartesian product and key-indexed data.
Free modules naturally give rise to \emph{polysets}, which generalise multisets and facilitate expressing database queries 
as multilinear maps with asymptotically efficient evaluation on polyset constructors.
We introduce \emph{compact maps} as a way of representing infinite (poly)sets constructible from an infinite base set and its elements by addition and subtraction. We show how natural joins generalise to algebraic joins, 
while intersection is implemented by a novel algorithm on \emph{nested} compact maps that carefully avoids visiting parts of the input that do not contribute to the eventual output. 
Our algebraic framework 
leads to a
\emph{worst-case optimal} evaluation of cyclic relational queries, which is known to be impossible using textbook query optimisers that operate on 
lists of records only.

\end{abstract}

\section{Introduction}

Query languages, like any other programming language, should be
\begin{itemize}
  \item \emph{efficient}, in that they should produce query results as
  efficiently as possible in terms of time, space and energy consumed; 
  \item \emph{expressive}, in that they should admit expressing solutions to many problems; and
  \item \emph{reasonable}, in that they should provide reasoning principles about a query's semantics to support compositional checking of functional and security properties, correctness of transformations and optimisations, query synthesis and more.
\end{itemize}
These aspects are sometimes presented as irreconcilable, suggesting, for example, that expressivity and reasonableness necessarily come at the cost of efficiency; or they may be studied in isolation of each other (expressiveness and correctness ``modulo'' efficiency, or efficient query engine implementation with specialised functionality ``modulo'' concerns what this does to the validity of reasoning principles in the core language).  We propose that efficiency, expressiveness and reasoning can mutually support each other via \emph{structure (algebraic and categorical)}.


Intuitively, our approach consists of identifying purely algebraic operations that can be implemented by symbolic simplification that exploits their algebraic properties for run-time efficiency, and factoring expensive exception and equality checking into separate operations that are invoked explicitly (e.g. checking whether a set is empty before deleting from it or whether two elements are equal or not) \emph{only when} this is required, rather than melding the two operations together into a single operation. To give meaning to the result of an exception-less deletion operation we introduce a data model that encompasses and goes beyond sets and multisets: \emph{polysets}, whose elements carry any integral multiplicity, not only $1$ (sets) or positive integers (multisets). Indeed, multiplicities can be replaced by arbitrary rings and even algebras.


We recognise finite polysets as the elements of the free module generated by a (usually infinite) set $X$ over $\mathbb{Z}$. 
More generally, we show that \emph{modules} over a commutative ring permit a variety of constructions that capture the core operations in query processing, including aggregation. They are all characterised by \emph{universal properties}, which not only justify singling them out as natural primitive operations, but also provide algebraic rewrite rules for run-time efficiency. Overall, our approach of building terms through free structures and then interpreting them using their universal properties is not unlike that taken by algebraic effects (see, e.g.,~\cite{Pretnar2015}).


Our approach is based on using free modules to represent generalised relations, biproducts for records, copowers for finite maps, tensor products for relational Cartesian products, and compact maps for infinite collections with wildcards. All of these constructions use arbitrary sets and modules to form new modules. We show that operations such as selection, projection, union, and intersection (and more) are not only very natural to express and reason about, but also lead us to highly efficient implementations of operations such as relational Cartesian products and, most significantly, arbitrary natural and outer joins.  In the case of joins, the asymptotic efficiency attained is not even achievable using conventional relational query optimisation technology. 


Although universal properties characterise these constructions uniquely \emph{extensionally}, some \emph{data structure representations} are better than others. A particular strength of our approach is how it uses prolific symbolic operator representations at run time to avoid unnecessary costly normalisation to a standard data structure, unless a particular context makes it absolutely necessary. For example, representing a tensor product symbolically gives a quadratic space compression and speed-up compared to normalising it to a list of atomic pairs; representing scalar multiplication symbolically eliminates iterated adding; representing additions symbolically facilitates operating on the summands independently, without first turning the tree of additions into a list of leaves.  More generally, this also applies to scalar multiplication so that the implementation of a linear map boils down to \emph{folding}, the composition of mapping the leaf elements to the target module and interpreting symbolic scalar multiplication and addition in the target module (which may also use symbolic scalar multiplication and addition).

Efficient evaluation hinges on the algebraic operations and their properties in a crucial way.  While lazy evaluation only delays standard evaluation, but does not avoid normalisation to a standard data structure, unnormalised representations using symbolic constructors let our evaluator invoke specialised evaluation rules depending on the particular way data are used; for example, computing the weight $\weight$ of a polyset (corresponding to the cardinality of a multiset) contains the clauses
$\weight(s_1 \otimes s_2) = \weight s_1 \cdot \weight s_2$ and $\weight(t_1 + t_2) = \weight t_1 + \weight t_2$. In particular, the tensor product $\otimes$ need not first be multiplied out to a quadratically bigger set of pairs of elements from $s_1$ and $s_2$.  Since the output of a join consists of a data dependent number of tensor products, this is not straightforwardly achievable by static preprocessing of a query.

Selective simplification is explicitly indicated and forced by efficient implementations of explicit natural isomorphisms to modules that correspond to data structures for efficient associative access.  This is where tensor products, copowers/finite maps and in particular compact maps, an algebraic generalisation of maps with a non-zero default value, come into play.


\paragraph{Implementation.}
These techniques are very general and can be applied and implemented in many ways, including as a library in a functional or logic language, a design pattern, or as a domain-specific language in its own right.  In this paper, we include an implementation in Haskell using type families to achieve both extensibility and efficiency.  All of the examples in this paper can be found in the Haskell source code. 

The main contributions of this paper are:

\begin{itemize}
\item An algebraic framework, based on universal constructions of modules and linear maps, into which generalised databases and queries can not only be interpreted, but evaluated highly efficiently.
\item An algebraic product,
formally a (weighted, unital) commutative algebra,
to express intersection on values and, via embedding of finite collections into the corresponding compact collections, of natural joins,
which in this case turn out to yield \emph{worst-case optimal} query evaluation.  This is illustrated for triangle queries, the canonical example of cyclic queries, which have only asymptotically suboptimal query implementations via classical relational query optimisation.  (The proof of worst-case optimality for algebraic joins, which include and extend relational joins, is beyond the scope of this paper.)
\item Efficient evaluation of expressions (queries) that leverages generic trie implementations of compact maps for efficient lookup, carefully scheduled enumeration for products (intersection and thus join), and aggressively exploiting algebraic simplifications.  In particular, all relevant operations of and on modules 
are purely symbolic when constructed, and only simplified according to the \emph{specific} context in which they are used.
\item An implementation of all of the above in Haskell.
\end{itemize}

This paper is structured as follows.
Section~\ref{prelim} defines the relevant algebraic structures, shows how to build modules and lists the most important isomorphisms.
Section~\ref{querylang} shows how linear maps can be used as an alternative to relational algebra.
Section~\ref{algeval} explains the simple, yet efficient approach to simplifying module terms.

\section{Mathematical Preliminaries}
\label{prelim}
We assume familiarity with fundamental (linear) algebraic structures including rings, modules, and (bi)linear maps, and refer the unfamiliar reader to the appendix. We recall here the definition of a (weighted, unital, commutative) algebra.

\begin{definition}[Algebra]
\label{algebra}
A \emph{commutative algebra} $S$ over commutative ring $\K$ is a commutative ring which is also a module over $\K$.
The addition and multiplication operations in the ring $S$ must coincide with those of the modules structure in their common domain.
When the algebra multiplication has a unit $e$ satisfying $e \cdot v = v \cdot e = v$ for all $v$ of $S$, we say that the algebra is \emph{unital}.
If additionally there is a linear map $\weight : S \to \K$ then $S$ is a \emph{weighted algebra}.
\end{definition}

For an element $x$ of a weighted algebra $\weight x$ is pronounced the \emph{weight} of $x$. As a proviso, we will henceforth leave out ``commutative''; all of our rings and algebras will be commutative.

\subsection{Meet the Modules}
We now present various useful ways of constructing modules.
The presentation is guided by category theory, but we will not introduce it explicitly.
Rather, our focus is on \emph{universal properties}, a system for abstractly defining objects as the most general solution given some condition.

\paragraph{Trivial modules.}
The simplest module is the zero module $\zerok$ consisting of just a single $0$ element.
All operations are trivial and determined by the axioms.
It satisfies the following universal property: for all modules $U$ there is a unique linear map $0 : \zerok \to U$, and a unique linear map $0 : U \to \zerok$.

The first interesting example of a module is the ring $\K$ itself with operations inherited from the ring structure.
This is called the \emph{scalar} module.
It also has a weighted algebra structure with multiplication inherited from the ring and weight given by the identity map.
The scalars satisfy the universal property that for any module $U$ and element $u : U$ there exists a unique linear map $f : \K \to U$ such that $f(1) = u$.

Intuitively the property states that any linear map from $\K$ is completely determined by what it does to $1$.
We are therefore justified in defining linear maps by pattern matching on $1$, for example defining some $f : \K \to \K$ by $f(1) = 42$. 
By linearity $f(r) = f(r \cdot 1) = r \cdot f(1) = r \cdot 42$ for any $r$, so we have exactly the information we need to determine the value of $f$ at any point.

\paragraph{Free modules.}
The \emph{free module} over a set $A$, denoted $\freek{A}$, is the module generated by elements of $A$.
By \emph{generated} we mean that for every element $a : A$ there is an element $\inj{a} : \freek{A}$ such that the set $\{\inj{a}\}_{a \in A}$ of all such elements form an orthonormal basis for $\freek{A}$.
Furthermore, $\freek{A}$ contains $0$ and is closed under addition and scalar multiplication.
For instance, there are elements like $3 \cdot \inj{a} + 5 \cdot \inj{b}$.
Two elements of $\freek{A}$ are equal if and only if they are forced to be equal by the module axioms and properties of $\K$.
Hence, $3 \cdot \inj{a} + 5 \cdot \inj{b} = 5 \cdot \inj{b} + 3 \cdot \inj{a}$ by commutativity, but $\inj{a} \neq 0$ (whenever $1 \neq 0$ in $\K$) since there is no way to show $\inj{a} = 0$ from the axioms alone (a formal argument for this is surprisingly tricky, though).
In particular $\inj{\cdot}$ is injective so $\inj{a} = \inj{b}$ implies $a = b$.

An element of $\freek{A}$ is best thought of as a generalised finite multiset.
Any such element can be written uniquely as a basis expansion $\sum_{a : A} (r_a \cdot \inj{a})$ where the number of non-zero $r_a$ is finite.
Depending on the nature of $\K$ there is a different interpretation of what `generalised finite multiset' means.

\begin{itemize}
\item When $\K$ is $\FinField{2}$ elements $\freek{A}$ are 
finite sets.
A set like $\s{a, b, c}$ is written as $\inj{a} + \inj{b} + \inj{c}$.
\item When $\K$ is $\Int$, elements of $\freek{A}$ are finite \emph{polysets}.
A polyset like $\s{a^3, b^{-2}, c^5}$ is written as $3 \inj{a} - 2 \inj{b} + 5 \inj{c}$.
\item When $\K$ is $\Real$ elements of $\freek{A}$ are generalised finite fuzzy sets, whose membership function is not limited to $[0,1]$.
A fuzzy set like $\s{a/0.3, b/0.2, c/0.5}$ is written as $0.3 \inj{a} + 0.2 \inj{b} + 0.5 \inj{c}$. 
\end{itemize}
The free module $\freek{A}$ satisfies the following universal property: for any module $V$ and function $f : A \to V$ there is a unique linear map $\ext{f} : \freek{A} \to V$ such that $\ext{f} \comp \inj{\cdot} = f$.
In essence, to define a \emph{linear map} out of $\freek{A}$ it suffices to identify the target module and define a \emph{map} out of $A$, and this map can be chosen \emph{freely}.
We can think of this as definition by pattern matching, and write linear maps like
\begin{eqnarray*}
g &:& \freek{\Str} \to \freek{\Str} \\
g(\inj{s}) &=& \inj{\mathbf{reverse}(s)}
\end{eqnarray*}
In this case $g = \ext{f}$ where $f(s) = \inj{\mathbf{reverse}(s)}$.
Note that due to linearity we also get the equations
\begin{align*}
g(0) &= 0 & g(x + y) &= g(x) + g(y) & g(r \cdot x) &= r \cdot g(x)
\end{align*}
but we do not need to handle these cases as they are forced by the condition of linearity.
Thus, we get to treat $\freek{A}$ as an inductive type and `pretend' that it only contains elements of $A$, even though it does contain many more elements than that.

Finally, we are going to equip $\freek{A}$ with a bilinear operator and a weight function to make it into a weighted algebra.
There is more than one possible choice, but only one makes sense for our purposes:
\begin{eqnarray*}
\inj{a} \cdot \inj{a} &=& \inj{a} \\
\inj{a} \cdot \inj{b} &=& 0 \quad \text{(for $a \neq b$)} \\
\weight \inj{a} &=& 1
\end{eqnarray*}
Note that we are defining multiplication by pattern matching in each argument, implicitly appealing to the universal property of $\freek{A}$ twice.
This operation can be thought of as a variant of the Kronecker delta function:
If the arguments $a, b$ are equal, we return that unique value as a singleton; if they are unequal, we `fail' by returning $0$.
When applied to sets it computes the set intersection; for multisets and polysets, however, the multiplicities of common elements are multiplied. 
For instance, for 
\[
(3 \inj{a} + 2 \inj{b} + 5 \inj{c}) \cdot (7 \inj{b} +  4 \inj{c} + 2 \inj{d}) = (2 \cdot 7) \inj{b} + (5 \cdot 4) \inj{c} = 14 \inj{b} + 20 \inj{c}
\]
The missing component in $\freek{A}$ being a \emph{unital} weighted algebra is the unit, an element $1$ such that $1 \cdot x = x = x \cdot 1$.
If $A$ is finite we can take $1 = \sum_{a : A} \inj{a}$, but for infinite $A$ this sum is not well-defined.
Thus, we now proceed to show how $\freek{A}$ can be extended to account for this deficiency, which in turn paves the way to efficient representation of certain infinite sets and---eventually---efficient algebraic join computations.

\paragraph{Compact free.}
We have seen that the free module over a set $A$ does an excellent job of representing finite subsets of $A$.
However, it lacks the ability to represent complements and in particular there is no general way to represent the subset containing every inhabitant of $A$.
Algebraically, the multiplication on $\freek{A}$ does not have a unit element for infinite $A$.

To rectify these deficiencies we introduce the \emph{compact free module}, $\cfreek{A}$, constructed by taking the free module $\freek{A}$ and adjoining a distinct element $1$.
We think of $1$ as \emph{symbolising} the potentially infinite sum $\sum_{a : A} \inj{a}$.
However, even when $A$ is finite, $1$ is by definition distinct from $\sum_{a : A} \inj{a}$.

The choice of never identifying $1$ with $\sum_{a : A} \inj{a}$ has several advantages.
We do not have to know whether $A$ is finite or infinite to decide if $1$ is linearly independent from the other generators.
It allows a compact symbolic representation of this sum when $A$ is finite but large.
And finally it gives us the following universal property: for any module $V$ together with a map $f : A \to V$ and an element $u : V$ there is a unique linear map $\ext{f} : \cfreek{A} \to V$ such that $\ext{f} \comp \inj{\cdot} = f$ and $\ext{f}(1) = u$.

In terms of pattern matching this amounts to having cases for $\inj{a}$ and $1$.
With this addition $\cfreek{A}$ has a unital weighted algebra structure, where multiplication of generators works just like for $\freek{A}$ and $1$ is the multiplicative unit.
Explicitly:
\begin{align*}
\inj{a} \cdot \inj{a} &= \inj{a} & 1 \cdot y &= y & \weight \inj{a} &= 1 \\
\inj{a} \cdot \inj{b} &= 0 \quad \text{(for $a \neq b$)} & x \cdot 1 &= x & \weight 1 &= 1
\end{align*}
This enables us to represent not just finite sets, but also \emph{cofinite} sets: the subsets of $A$ that contain all but a finite number of elements.
For example, the set $A \setminus \s{a, b}$ is written as $1 - (\inj{a} + \inj{b})$.
When we introduce tensor products later we will see how even more interesting subsets can be represented compactly in this manner.

\paragraph{Biproduct.}
The \emph{biproduct} of modules $U$ and $V$ is a module $U \oplus V$ consisting of pairs of elements from $U$ and $V$ with operations defined pointwise.
For instance given $(u_1, v_1), (u_2, v_2) : U \oplus V$ we define
\[
(u_1, v_1) + (u_2, v_2) = (u_1 + u_2, v_1 + v_2)
\]
Categorically speaking, the biproduct is both a binary product and a binary coproduct such that injection and projection is compatible.
For details see Appendix~\ref{appendix:biproduct}.

\paragraph{Finite map.}
Suppose we are given a set $A$ and a module $U$.
The \emph{finite map} module $A \fmap U$ consists of maps $A \to U$ with \emph{finite support}, i.e. maps which produce a non-zero value at finitely many elements of $A$.
It is the module generated by elements of the form $a \mapsto u$ where $a : A$ and $u : U$, subject to the requirement that $a \mapsto \cdot$ is a linear map for any $a$.
Any finite map can be written as $\sum_{a : A} (a \mapsto u_a)$ where $u_a = 0$ for all but finitely many $a$.

The finite map module satisfies the following universal property: for any module $V$ together with a family of maps $f_a : U \to V$ there exists a unique linear map $\casemap{f} : (A \fmap U) \to V$ such that $\casemap{f} \comp (a \mapsto \cdot) = f_a$ for all $a : A$.
This allows 
pattern matching similar to the free module.
For example:
\begin{eqnarray*}
f &:& (A \fmap \K) \to \freek{A} \\
f(a \mapsto 1) &=& \inj{a}
\end{eqnarray*}
Recall that the use of $1$ on the left-hand side is pattern matching on scalars.

Intuitively, we think of $A \fmap U$ as elements of $U$ indexed by $A$.
For instance, suppose we have a multiset of strings and we want to index it by string lengths.
The indexed space would be $\Nat \fmap \freek{\Str}$ with the indexing being done as follows.
\begin{eqnarray*}
\mathrm{indexbylength} &:& \freek{\Str} \to (\Nat \fmap \freek{\Str}) \\
\mathrm{indexbylength}(\inj{s}) &=& \mathbf{length}(s) \mapsto \inj{s}
\end{eqnarray*}
We can also go in the other direction and forget the index.
\begin{eqnarray*}
\mathbf{sum} &:& (A \fmap U) \to U \\
\mathbf{sum}(a \mapsto u) &=& u
\end{eqnarray*}
Like the free module the finite map module is a weighted algebra, but lacks a unit when the index set is infinite.
Multiplication and weight are defined by:
\begin{eqnarray*}
(a \mapsto u) \cdot (a \mapsto v) &=& a \mapsto (u \cdot v) \\
(a \mapsto u) \cdot (b \mapsto v) &=& 0 \quad \text{(for $a \neq b$)} \\
\weight (a \mapsto u) &=& \weight u
\end{eqnarray*}

\paragraph{Compact map.}
Finite maps, like free modules, do not possess a unit element when the index set is infinite.
More generally, they do not possess constant mappings that have the same (non-zero) value everywhere.
The solution is similar: adjoin distinct elements to account for this deficiency.

Thus, the \emph{compact map} module $A \cmap U$ is obtained by adding elements of the form $* \mapsto u$ for any $u : U$.
The $*$ represents a \emph{wildcard} which matches anything.
Intuitively $* \mapsto u$ symbolises $\sum_{a : A} (a \mapsto u)$ but is by definition always distinct, just like for the compact free module.

The compact map module satisfies the following universal property: for any module $V$ together with a family of maps $f_a : U \to V$ as well as a map $f_* : U \to V$ there exists a unique linear map $\casemap{f} : (A \cmap U) \to V$ such that $\casemap{f} \comp (a \mapsto \cdot) = f_a$ for all $a : A$ and $\casemap{f} \comp (* \mapsto \cdot) = f_*$.

The unital weighted algebra structure on compact maps is given as follows.
\begin{align*}
(a \mapsto u) \cdot (a \mapsto v) &= a \mapsto (u \cdot v) & 1_{A \cmap U} &= * \mapsto 1_U \\
(a \mapsto u) \cdot (b \mapsto v) &= 0 \quad \text{(for $a \neq b$)} \\
(a \mapsto u) \cdot (* \mapsto v) &= a \mapsto (u \cdot v) & \weight (a \mapsto u) &= \weight u \\
(* \mapsto u) \cdot (a \mapsto v) &= a \mapsto (u \cdot v) & \weight (* \mapsto u) &= \weight u \\
(* \mapsto u) \cdot (* \mapsto v) &= * \mapsto (u \cdot v)
\end{align*}
Lookup in a compact map works like for finite maps, except $*$ is also a valid key.
To see how this works in practice, suppose we have an element:
\[
x = (* \mapsto 2) + (a \mapsto 3) + (b \mapsto -2)
\]
Consider the following lookups:
\begin{align*}
x(*) &= 2 & x(a) &= 2 + 3 = 5 & x(b) &= 2 - 2 = 0 & x(c) &= 2
\end{align*}
It is evident that the $* \mapsto 2$ component of $x$ serves as a \emph{baseline} and a component like $a \mapsto 3$ determines how much the value at $a$ \emph{deviates} from that baseline.
In particular, the value at $b$ deviates by exactly the negative of the baseline value, the result being that $b$ is contained $0$ times in $x$ when viewed as a polyset.
This is in contrast to the more common construct of having finite maps with a \emph{default value} for keys not explicitly listed.  In particular, compact map addition is commutative: it does not matter in which order the maps are listed.

\paragraph{Tensor product.}
The \emph{tensor product} of modules $U$ and $V$ is a module $U \otimes V$ generated by elements of the form $u \otimes v$ with $u : U$ and $v : V$, subject to the requirement that $\otimes$ is bilinear, i.e. linear in each argument separately.
More explicitly, linearity in the first argument requires
\begin{align*}
0 \otimes v &= 0 & (u_1 + u_2) \otimes v &= u_1 \otimes v + u_2 \otimes v & (r \cdot u) \otimes v &= r \cdot (u \otimes v)
\end{align*}
Linearity in the second argument is analogous.
In general, any element of the tensor product can be written as a sum of $\otimes$-pairs, i.e. $\sum_i (u_i \otimes v_i)$.
Compare this with the biproduct, which is also generated by pairs of elements.
An element $\sum_i (u_i, v_i)$ can always be reduced to $(\sum_i u_i, \sum_i v_i)$, since the biproduct pair constructor is linear in both arguments \emph{simultaneously}.
The tensor product is linear in each argument \emph{separately} and elements can only be simplified in some circumstances, such as
\[
u_1 \otimes v_1 + u_1 \otimes v_2 + u_2 \otimes v_1 + u_2 \otimes v_2 = (u_1 + u_2) \otimes (v_1 + v_2)
\]
This kind of simplification can often reduce the size of a term from quadratic to linear, but recognising such opportunities is not easy.
Consequently, when a term is in this kind of simplified, compact form we should not expand it unless absolutely necessary!

The tensor product satisfies the following universal property: for any module $W$ and bilinear map $f : U \times V \to W$ there exists a unique linear map $g : U \otimes V \to W$ such that $g \comp \otimes = f$.

The consequence is that 
any bilinear map can be written as a linear map from the tensor product.
It also justifies defining linear maps by pattern matching on $\otimes$, for example:
\begin{eqnarray*}
f &:& U \otimes V \to V \otimes U \\
f(u \otimes v) &=& v \otimes u
\end{eqnarray*}
This is only valid if the definition is linear in each argument separately.
In particular we cannot simply define a projection $\pi_1 : U \otimes V \to U$ as $\pi_1(u \otimes v) = u$, as that would not be linear in the second argument.
In this respect the tensor product is different from an ordinary product type.
If $V$ is equipped with a weighted algebra structure, however, we can define:
\begin{eqnarray*}
\pi_1 &:& U \otimes V \to U \\
\pi_1(u \otimes v) &=& \weight v \cdot u
\end{eqnarray*}
The weight serves to dispose of data, which is not possible for an arbitrary module in general.
Finally, we equip the tensor product with an algebra structure.
Suppose $U$ and $V$ are algebras.
All operations are defined pointwise:
\begin{align*}
&(u_1 \otimes v_1) \cdot (u_2 \otimes v_2) = (u_1 \cdot u_2) \otimes (v_1 \cdot v_2) \\
&1_{U \otimes V} = 1_U \otimes 1_V \qquad \weight (u \otimes v) = \weight u \cdot \weight v
\end{align*}
Note in particular how computing the weight of a tensor product simply reduces to computing the weight of each factor.
This saves us from having to expand $u \otimes v$ at all.
When $u$ and $v$ are themselves large sums this saves a considerable amount of work.

\subsection{Functors and isomorphisms}
All of the constructions we have seen are \emph{functors}, structure-preserving maps between categories.
Briefly, a functor acts not just on objects (sets, modules, etc.), but also on maps between objects.
For instance $\freek{A}$ is functorial in $A$, so given any map between sets $f : A \to B$ we have $\freek{f} : \freek{A} \to \freek{B}$.

Let $f$ be a map between sets and $\alpha, \beta$ be linear maps.
The functors act as follows:
\begin{align*}
\freek{f}(\inj{a}) &= \inj{f(a)} & 
\cfreek{f}(\inj{a}) &= \inj{f(a)} \\
\cfreek{f}(1) &= 1 &
(\alpha \otimes \beta)(u \otimes v) &= \alpha(u) \otimes \beta(v) \\
(f \cmap \alpha)(* \mapsto u) &= * \mapsto \alpha(u) &
(f \cmap \alpha)(a \mapsto u) &= f(a) \mapsto \alpha(u) \\
(f \fmap \alpha)(a \mapsto u) &= f(a) \mapsto \alpha(u)
\end{align*}
Generally, these actions can be derived mechanically and there is only one reasonable choice.
The simplicity is deceptive, though, and it is easy to overlook how much one gets for free with a categorical approach.
Perhaps the clearest example of this is $\otimes$.
It is usually called the Kronecker product, defined as a complicated block matrix expression and relegated to advanced linear algebra courses.
The functorial action, by contrast, could not be simpler.

\begin{table}[tb]
\footnotesize
\begin{align*}
  \freek{0} & \iso \zerok & \freek{1} & \iso \K \\
  0 \fmap U & \iso \zerok & 1 \fmap U & \iso U \\ 
  A \fmap \zerok & \iso \zerok & A \fmap \K & \iso \freek{A} \\ 
  \freek{A + B} & \iso \freek{A} \oplus \freek{B} & \freek{A \times B} & \iso 
  \freek{A} \otimes \freek{B} \\
  (A + B) \fmap U & \iso (A \fmap U) \oplus (B \fmap U) & (A \times B) \fmap U 
  & \iso A \fmap B \fmap U \\
  A \fmap (U \oplus V) & \iso (A \fmap U) \oplus (A \fmap V) & A \fmap (U 
  \otimes V) & \iso (A \fmap U) \otimes V \\
  A \fmap U & \iso \freek{A} \otimes U & A \cmap U & \iso \cfreek{A} \otimes 
  U\\ 
  \cfreek{A} & \iso \freek{A} \oplus \K & A \cmap U & \iso (A \fmap U) \oplus A
\end{align*}
\normalsize

\caption{A selection of natural isomorphisms}
\label{iso}
\end{table}

The module constructions we have presented are related in various ways.
More precisely, there are a number of \emph{natural isomorphisms}.
An isomorphism $\varphi : U \iso V$ is simply a linear map $\varphi : U \to V$ together with an inverse linear map $\varphi^{-1} : V \to U$.
Naturality can be stated precisely using category theory, but for our purposes it is sufficient to think of `natural' as `polymorphic'.

When defining an isomorphism $\varphi$ we 
write both directions simultaneously using the syntax $p \isodef q$ to mean $\varphi(p) = q$ and $p = \varphi^{-1}(q)$.
A selection of isomorphisms can be seen in Table~\ref{iso}.
We 
draw particular attention to the relationship between free modules and tensor products 
given by the 
isomorphism:
\begin{eqnarray*}
\freek{A \times B} &\iso& \freek{A} \otimes \freek{B} \\
\inj{(a, b)} &\isodef& \inj{a} \otimes \inj{b}
\end{eqnarray*}
The tensor product of free modules can itself be written as a free module.
However, the pattern matching notation belies the true cost of this conversion.
The typical element of $\freek{A} \otimes \freek{B}$ is a sum of terms like $(\sum_i r_i \inj{a_i}) \otimes (\sum_j s_j \inj{b_j})$ which converts to $\sum_i \sum_j r_i s_j \inj{(a_i, b_j)}$, a quadratic increase in size.
Converting back again yields $\sum_i \sum_j r_i s_j (\inj{a_i} \otimes \inj{b_j})$.
This is extensionally equal to the original term, but much larger.
Hence, passing through the free module is not free!
We will generally prefer to stay on the right side of this isomorphism, only converting when necessary. Fortunately, the isomorphisms shown thus far demonstrate that any polynomial type can be expressed as a module using $\zerok$, $\K$, $\oplus$ and $\otimes$.

\subsubsection{Free Modules Everywhere?}
At this point we have seen that any module constructed using any combination of $\zerok$, $\K$, $\oplus$, $\otimes$, $\freek{\cdot}$, $\cfreek{\cdot}$ and $\fmap$ is isomorphic to some free module.
Indeed, if $\K$ is a field---in which case modules are vector spaces---the classical mathematician would remark that by the Axiom of Choice every vector space is isomorphic to a free one.
Why do we not simply consider only free modules then?

Firstly, these isomorphisms only concern the module structure.
The algebra structure differs in many cases.
For instance, we saw that $\cfreek{A} \iso \freek{A} \oplus \K$ as modules, but certainly \emph{not} as algebras (if the right-hand side even \emph{has} an algebra structure, which is only the case when $A$ is finite).
Secondly, even if two modules are \emph{extensionally} equal, i.e. isomorphic, they need not be \emph{intensionally} equal.
The clearest example of this is $\freek{A \times B} \iso \freek{A} \otimes \freek{B}$ where the right-hand side can express certain large terms much more compactly and converting to the left-hand side generally yields asymptotically larger terms.

\section{Linear Algebra as a Query Language}
\label{querylang}
Relational algebra serves as the traditional formalism underpinning query languages.
We propose 
the theory of modules and the linear maps between them as an appealing generalised framework for expressing queries.
So far we have seen that linear algebra can express a more general class of sets than the usual kinds of sets and multisets employed in query languages; in particular, it is possible to have sets with negative multiplicities, cofinite sets and more.  
We now present a `Rosetta Stone' showing how the operations of relational algebra have corresponding linear maps on modules.

\paragraph{Selection}
In relational algebra \emph{selection} restricts a set of tuples to the subset satisfying a given predicate.
Suppose $P \subseteq A$. 
We define $\sigma_P : \freek{A} \to \freek{A}$ by
\begin{align*}
\sigma_P(\inj{a}) & = \inj{a} \quad \text{(when $a \in P$)} &
\sigma_P(\inj{a}) = 0 \quad \text{(when $a \notin P$)}
\end{align*}
The effect is that a generator $\inj{a}$ is preserved when $a \in P$ and eliminated otherwise.

\paragraph{Projection}
In relational algebra \emph{projection} selects a subset of attributes, throwing away those not in the designated subset.
Note that due to set semantics this may cause values to collapse, making the resulting set smaller.
For example projecting the $A$ component of $\s{(A : \mathtt{foo}, B : 1), (A : \mathtt{foo}, B : 2), (A : \mathtt{bar}, B : 3)}$ yields $\s{(A : \mathtt{foo}), (A : \mathtt{bar})}$.
For a tensor product $\freek{A} \otimes \freek{B}$ we define the two projections as follows.
\begin{align*}
& \pi_1 : \freek{A} \otimes \freek{B} \to \freek{A} & \pi_1 (x \otimes y) = \weight y \cdot x \\
& \pi_2 : \freek{A} \otimes \freek{B} \to \freek{B} & \pi_2 (x \otimes y) = \weight x \cdot y
\end{align*}
More complicated projections can be constructed using these two projections, the identity map, and the functorial action of $\otimes$.
Note that, unlike relational algebra, all such projections preserve multiplicities.

\paragraph{Renaming.}
In relational algebra \emph{renaming} changes the names of attributes.
This makes sure everything is only done ``up to choice of names'', but it is also sometimes necessary in order to apply a natural join.

We do not use named attributes, but a similar concern arises with regards to the order and parenthesisation of attributes.
To this end we have two natural isomorphisms, the \emph{associator} and the \emph{commutator}:
\begin{align*}
\alpha &: U \otimes (V \otimes W) \iso (U \otimes V) \otimes W & \beta &: U \otimes V \iso V \otimes U
\end{align*}
Combining these isomorphisms with the functorial action of $\otimes$, we can rearrange arbitrarily as needed. 

\paragraph{Union and intersection.}
Union in relational algebra is just the usual union of sets.
We define union as simply $+$.
In general this form of union keeps track of multiplicities so for instance
$
(\inj{a} + \inj{b}) + (\inj{b} + \inj{c}) = \inj{a} + 2 \inj{b} + \inj{c}
$.

In relational algebra \emph{intersection} is typically not mentioned explicitly, but it arises as a special case of join when the two relations have the same set of attributes.
In our approach intersection is the primitive upon which join is built.
It is simply the product operation from the algebra structure.
If $x, y : \freek{A}$ then $x y : \freek{A}$ is their intersection.
Recall that the algebra product is a bilinear operator so multiplicities are multiplied.
For instance
$
(2 \inj{a} + 3 \inj{b})(5 \inj{b} + 7 \inj{c}) = (3 \cdot 5) \inj{b} = 15 \inj{b}
$.

\paragraph{Cartesian product.}
Cartesian product in relational algebra is also the usual notion from set theory, and
is traditionally viewed as an expensive operation that is best avoided if possible.
Our version of the Cartesian product is the tensor product.
We view it not as an operation, but as a symbolic term.
In particular, we are perfectly comfortable writing down terms like
\[
t = (\inj{a_1} + \cdots + \inj{a_m}) \otimes (\inj{b_1} + \cdots + \inj{b_n})
\]
\emph{without} insisting that this be expanded to a canonical form as soon as possible.
Such an expansion generally incurs a quadratic blow-up in expression size.
Depending on what happens to $t$ later we might never have to expand it.
For instance, $\pi_2(t)$ can be computed as
\[
\pi_2(t) = \weight(\inj{a_1} + \cdots + \inj{a_m}) \cdot (\inj{b_1} + \cdots + \inj{b_n}) = m \cdot (\inj{b_1} + \cdots + \inj{b_n})
\]
The amount of work done was linear in the size of the symbolic term, and sublinear in the size of the hypothetically expanded form.
Also note that even this result is not yet in canonical form---there is no need to distribute the scalar multiplication prematurely.

Static analysis will typically recognise opportunities like a projection immediately applied to a Cartesian product and do appropriate optimisation.
By contrast, our approach does it at run time.
It does not rely on a sufficiently clever analysis, it guarantees that products are not expanded until necessary and it even allows terms to be stored more efficiently in data structures between operations.

\paragraph{Natural join}
In relational algebra the \emph{natural join} of two relations is constructed by taking their Cartesian product and keeping only the tuples where both sides agree about the values of shared attributes.

For example, consider the relations
\begin{align*}
x = \{&(A : a, B : 1), & y = \{&(B : 2, C : p) \\
&(A : b, B : 2), & & (B : 3, C : q), \\
&(A : c, B : 3) & &(B : 4, C : r) \}
\end{align*}
where the notation $(A : a, B : 1)$ denotes a tuple with a value of $a$ for attribute $A$ and $1$ for attribute $B$.
Their join is
\[
x \Join y = \s{(A : b, B : 2, C : p), (A : c, B : 3, C : q)}
\]
In our system these two relations would be 
represented as the vectors $x$ and $y$ given by
\begin{align*}
x &: \freek{\Str} \otimes \freek{\Int} & x &= \inj{a} \otimes \inj{1} + \inj{b} \otimes \inj{2} + \inj{c} \otimes \inj{3} \\
y &: \freek{\Int} \otimes \freek{\Str} & y &= \inj{2} \otimes \inj{p} + \inj{3} \otimes \inj{q} + \inj{4} \otimes \inj{r} \enspace.
\end{align*}
To compute their join we first have to inject them into a common module.
This is done by going from $\freek{\cdot}$ to $\cfreek{\cdot}$ and adding $1$'s as necessary.
\begin{align*}
x' &: \cfreek{\Str} \otimes \cfreek{\Int} \otimes \cfreek{\Str} & x' &= \inj{a} \otimes \inj{1} \otimes 1 + \inj{b} \otimes \inj{2} \otimes 1 + \inj{c} \otimes \inj{3} \otimes 1 \\
y' &: \cfreek{\Str} \otimes \cfreek{\Int} \otimes \cfreek{\Str} & y' &= 1 \otimes \inj{2} \otimes \inj{p} + 1 \otimes \inj{3} \otimes \inj{q} + 1 \otimes \inj{4} \otimes \inj{r}
\end{align*}
The join of two elements of the same module is simply their intersection, which is given by multiplication.
A naive approach to simplification is to apply distributivity bluntly and then simplify using identities for tensor products (see Appendix~\ref{appendix:naiveintersection} for details).
In the end we get the simplified form:
\[
\inj{b} \otimes \inj{2} \otimes \inj{p} + \inj{c} \otimes \inj{3} \otimes \inj{q}
\]
This method of simplification is wasteful as we are expanding everything using distributivity only to have most components turn out to be $0$.
We shall later see that there is a much more efficient approach to simplifying expressions that in particular can solve basic joins like this one in linear time.

\paragraph{Outer join}
In relational algebra the \emph{outer join} is similar to the natural join.
The difference is that tuples from either input which would not occur anywhere in the output are included anyway.
The missing attributes are populated with a $\Null$ value.

Consider the previous example of a natural join.
The result of the outer join would be as follows.
\small
\[
\s{(A : a, B : 1, C : \Null), (A : b, B : 2, C : p), (A : c, B : 3, C : q), (A : \Null, B : 4, C : r)}
\]
\normalsize
There are also \emph{left outer join} and \emph{right outer join} operations, which only include extra tuples from the left and right input respectively.

In our system we use wildcards to achieve a similar effect.
Note that though the wildcard bears superficial similarity to $\Null$, it is in fact the exact opposite: $\Null$ is a special value that agrees with \emph{no value} (including itself), the wildcard is a special value that agrees with \emph{every value}.
It could be argued that the main reason $\Null$ is used to fill missing values in relational algebra is that it is the only somewhat applicable tool in the relational toolbox.

Recall that to compute the natural join of $x$ and $y$ we first embed them by adding wildcards to get $x'$ and $y'$.
The left outer join is then given by $x' \cdot (y' + 1)$, the right outer join by $(x' + 1) \cdot y'$, and the outer join by $(x' + 1) \cdot (y' + 1)$.
By distributivity this is the same as $x' \cdot y' + x'$, $x' \cdot y' + y'$ and $x' \cdot y' + x' + y' + 1$ respectively.
One way to think about an expression like $x' \cdot (y' + 1)$ is that the added $1$ ensures that every component of $x'$ matches with \emph{something}.

For the outer join one might want to use $(x' + 1) \cdot (y' + 1) - 1$ instead to avoid the last factor of $1$.
However, our proposed definition generalises more neatly to $n$-ary outer joins, which can be written simply as $(x_1 + 1) \cdots (x_n + 1)$.

\paragraph{Aggregation}
In (extensions of) relational algebra \emph{aggregation} computes values like sum or maximum over an attribute.
For instance, given the relation
$
\s{(A : p, B : 2), (A : p, B : 3), (A : q, B : 4)}
$
aggregating $B$ by sum would yield
$
\s{(A : p, B : 5), (A : q, B : 4)}
$,
while aggregating $B$ by maximum would yield
$
\s{(A : p, B : 3), (A : q, B : 4)}
$.

We take a different view of aggregation, as something that happens implicitly due to module semantics.
This was evident when discussing projection, where the discarded attributes are automatically counted with multiplicities.
The specifics of aggregation depends on the nature of the ring $\K$.

Take the example above and represent it as follows:
$
2 \inj{p} + 3 \inj{p} + 4 \inj{q} = (2 + 3) \inj{p} + 4 \inj{q}
$. 
If $\K$ is $\Int$ with ordinary arithmetic this reduces to $5 \inj{p} + 4 \inj{q}$.
Note how a tuple such as $(A : p, B : 2)$ was represented as $2 \inj{p}$ and not as $\inj{p} \otimes \inj{2}$.
We \emph{moved} the attribute into the ring.

Not all aggregations can be expressed directly as a ring structure, though.
For instance, integers extended with infinity and equipped with minimum and maximum operations only form a \emph{semiring} since negation is impossible.
Our approach can easily work with semirings as well, but as we shall see shortly negation plays an important role and should not be given up so easily.

Instead, we express the aggregation using nested free modules to group data.
Each grouping is an element of $\free{\FinField{2}}{A}$, i.e. an ordinary finite set.
The example above would be represented as follows:
$
\inj{p} \otimes \inj{\inj{2} + \inj{3}} + \inj{q} \otimes \inj{\inj{4}}
$. 
Non-linear aggregations like minimum are modelled as functions (\emph{not} linear maps) $\min : \free{\FinField{2}}{\Int} \to \IntInf$ with
$
\min(\inj{a_1} + \cdots + \inj{a_n}) = \min\s{a_1, \ldots, a_n}
$. 
Here $\IntInf$ is $\Int$ with $\infty$ adjoined so that $\min\s{}$ is well-defined.
Aggregating by minimum on the second attribute we get:
\[
(\id{\freek{\Str}} \otimes \freek{\min})(\inj{p} \otimes \inj{\inj{2} + \inj{3}} + \inj{q} \otimes \inj{\inj{4}}) = \inj{p} \otimes \inj{2} + \inj{q} \otimes \inj{4}
\]
In general, nested free modules allow any non-linear function to be included in an otherwise linear query.
Besides aggregations, this includes operations like turning negative multiplicities into zeroes.

These considerations demonstrate that linear maps are exactly the maps that can be easily adapted to a distributed setting.
The non-linear maps, on the other hand, depend on having the entire dataset at hand and subjected to at least some degree of simplification, which implies that synchronisation is necessary.

\paragraph{Domain computations.}
Relational algebra proper does not provide a way to transform data, but in practical realisations this is usually possible.
For instance, given the relation $\s{(A \mapsto \mathrm{foo}), (A \mapsto \mathrm{bar})}$ we might transform it using a function $\mathbf{upper}$ that maps strings to uppercase, getting the result $\s{(A \mapsto \mathrm{FOO}), (A \mapsto \mathrm{BAR})}$.
In our system this is achieved by the functorial action of $\freek{\cdot}$.
In particular given $\mathbf{upper} : \Str \to \Str$ we have $\freek{\mathbf{upper}} : \freek{\Str} \to \freek{\Str}$.
We can therefore apply it to an element as follows:
$
\freek{\mathbf{upper}}(\inj{\mathrm{foo}} + \inj{\mathrm{bar}}) = \inj{\mathbf{upper}(\mathrm{foo})} + \inj{\mathbf{upper}(\mathrm{bar})} = \inj{\mathrm{FOO}} + \inj{\mathrm{BAR}}
$.

\paragraph{Insert and delete.}
For persistent relations there is the question of how to do updates.
Relational algebra is based on set theory, so updates can be done using set union and difference.

In our system all updates are done by addition.
Deleting an element amounts to adding its negative.
For example, say the database contains $\inj{a} + \inj{b}$ and we want to add $c$ and delete $b$.
This update is computed as
$
(\inj{a} + \inj{b}) + (\inj{c} - \inj{b}) = \inj{a} + \inj{c}
$.

Note that there is no conceptual distinction between \emph{databases} and \emph{database updates}, since deletions are simply represented as negative elements.
Furthermore, since $+$ is commutative the order of updates does not matter.
For instance, suppose the database only contains $\inj{a}$ and we run the update above:
$
\inj{a} + (\inj{c} - \inj{b}) = \inj{a} - \inj{b} + \inj{c}
$.
This leaves a database with a negative occurrence of $b$.
If we then insert $b$ afterwards we end up with $\inj{a} + \inj{c}$ again.
In a sense this is the easiest possible conflict resolution strategy: just accept the data from every update and add it all together!
Of course, there are times when we might want to ensure that the database does not contain negative multiplicities by, say, setting them all to $0$ using an explicit operation. 
In that case---and that case only---we separate updates into before and after that given 
point.

\section{Algebraic Evaluation}
\label{algeval}
We now turn to the question of evaluation.
More precisely, the question of \emph{simplifying} module terms.
What kind of simplified form do we have in mind; is simplification even necessary?
That depends on the context of use, but generally a simplified form involves some of the following:
\begin{itemize}
\item No zeroes in additions or multiplications, and no multiplications except $r \cdot \inj{a}$ where $r : \K$.
\item No repeated generators, e.g. $\inj{a} + \inj{b} + \inj{a}$ should be $2 \inj{a} + \inj{b}$.
\item No sums of biproducts, e.g. $(u_1, v_1) + (u_2, v_2)$ should be $(u_1 + u_2, v_1 + v_2)$.
\item No tensor products where either factor is a sum.
\end{itemize}
The kind of observation we wish to make dictates which 
requirements are necessary.
For example, the most elementary observation we can make is to ask if a term is equal to $0$.
In query terms this corresponds to the question of satisfiability.
We certainly need to avoid repeated generators as otherwise they might happen to cancel out, e.g. $\inj{a} - 2\inj{a} + \inj{a}$ is a non-trivial representation of $0$.
On the other hand, if the term is a tensor product we do not need to expand it, since $u \otimes v$ is $0$ precisely when either factor is $0$.

In discussing 
term simplification it is important to distinguish between \emph{intensional} and \emph{extensional} equality.
Two terms are intensionally equal when they are written the same way (modulo renaming).
They are extensionally equal when they can be shown to be equal using the presented algebraic identities.
For example, $0 + 0 + 0 + 0$ and $0$ are extensionally equal which can be argued using the identity $x + 0 = x$ repeatedly.
However, they are not intensionally equal and indeed we would consider the latter a more compact version of the former.

\subsection{Finite maps}
Before dealing with simplification of arbitrary terms we consider finite maps in particular.
They have a great deal of structure to exploit and also play a vital r\^ole in making simplification efficient.

Suppose we have some term $x : A \fmap U$ and \emph{assume that we already have a method for simplifying terms from $U$}.
To simplify $x$ we proceed depending on the nature of $A$.
Recall that we have the following isomorphisms at our disposal:
\begin{align*}
\fmapemptyiso : 0 \fmap U &\iso \zerok & 
\fmapsumiso : (A + B) \fmap U &\iso (A \fmap U) \oplus (B \fmap U) \\
\fmapunitiso : 1 \fmap U &\iso U &
\fmapprodiso : (A \times B) \fmap U &\iso A \fmap B \fmap U
\end{align*}
The idea is that we wrap the term in the appropriate isomorphism.
To start, suppose $A = 1$ so $x : 1 \fmap U$.
We can then write $x$ as $\fmapunitiso^{-1}(\fmapunitiso(x))$.
The term $\fmapunitiso(x)$ is then simplified to get some term $y : U$ and the simplified form of $x$ becomes $\fmapunitiso^{-1}(y)$.

Now suppose $A = A_1 + A_2$ so $x : (A_1 + A_2) \fmap U$.
We write $x$ as $\fmapsumiso^{-1}(\fmapsumiso(x))$.
The term $\fmapsumiso(x)$ reduces to some term $(y_1, y_2) : (A_1 \fmap U) \oplus (A_2 \fmap U)$ (a biproduct can always be reduced to a pair by simple component-wise addition).
At this point $y_1$ and $y_2$ are finite maps with index sets $A_1$ and $A_2$ respectively, so we apply the procedure recursively to get simplified forms $z_1$ and $z_2$.
The simplified form of $x$ then becomes $\fmapsumiso^{-1}(z_1, z_2)$.

Finally, suppose $A = A_1 \times A_2$ so $x : (A_1 \times A_2) \fmap U$.
In this case $\fmapprodiso(x)$ has type $A_1 \fmap A_2 \fmap U$.
Recursively we have a procedure to simplify terms from $A_2 \fmap U$; and given this we also get a procedure to simplify $\fmapprodiso(x)$ to get some term $y$.
The simplified form of $x$ becomes $\fmapprodiso^{-1}(y)$.

What about sets that are not sums or products?
Recursively defined sets do not pose any extra challenge; we omit the details.
Function spaces, i.e. $A = A_1 \to A_2$, are not amenable to simplification in general and we exclude them from consideration.
Finally, there are primitives like $n$-bit integers.
These can be handled using specific methods.
In the case of finite precision integers an efficient solution is to represent finite maps as Patricia tries.
For strings a good choice would be a radix trie.

An example of this simplication procedure in action can be found in Appendix~\ref{appendix:simplification-example}.

\subsection{Simplification}
We now sketch the general approach to simplification of terms.
Suppose we have terms $u_1, \ldots, u_n : \cfreek{A_1} \otimes \cdots \otimes \cfreek{A_m}$ and we want to simplify the product $u_1 \cdots u_n$.
The first step is to exploit the isomorphism $A \cmap U \iso \cfreek{A} \otimes U$ for each $A_i$ to get corresponding simplified terms $v_1, \ldots, v_n : A_1 \cmap \cdots \cmap A_m \cmap \K$.
Simplification then proceeds by dealing with one level of this nested map structure at a time.
One of the terms will act as enumerator and the others will act as filters.
Find the $v_i$ whose expression as a sum contains the fewest components.
Without loss of generality assume that this is $v_1$.
For ease of presentation assume also that $A_1$ is a primitive set so we can write the simplified form of $v_1$ as $(\sum_{1 \le i \le k} a_i \mapsto v_{1,i}) + (* \mapsto v_{1,*})$.
We apply distributivity to move everything under this summation, and we turn the intersections with $v_2 \cdots v_n$ into lookups.
\begin{eqnarray*}
v_1 v_2 \cdots v_n &=& ((\ssum_{1 \le i \le k} a_i \mapsto v_{1,i}) + (* \mapsto v_{1,*})) v_2 \cdots v_n \\
&=& (\ssum_{1 \le i \le k} (a_i \mapsto v_{1,i}) v_2 \cdots v_n) + (* \mapsto v_{1,*}) v_2 \cdots v_n \\
&=& (\ssum_{1 \le i \le k} a_i \mapsto v_{1,i} v_2(a_i) \cdots v_n(a_i)) + (* \mapsto v_{1,*} v_2(*) \cdots v_n(*))
\end{eqnarray*}
At this point there are intersection subterms of type $A_2 \cmap \cdots \cmap A_n \cmap \K$, so we can apply the procedure recursively.
In the base case when $m = 0$ the problem is reduced to simple multiplication in $\K$.
Finally, the resulting iterated compact map is turned back into a tensor product of compact free modules.

The essence of this approach is that we deal with one attribute at a time and simplify with regards to that attribute across the entire term.
Distributivity is only applied as necessary to proceed.
This general strategy works for any term, not just multiplications.

\subsection{Worst-case optimality}
The particular case of a multiplication of sums correspond to \emph{conjunctive queries} in relational algebra.
Conjunctive queries have a simple structure, but are difficult to evaluate efficiently.
The main obstacle to efficiency is what is known as cyclic queries.
The simplest one is the triangle query, which given attributes $A$, $B$ and $C$ asks to compute the join of $x : \freek{A} \otimes \freek{B}$, $y : \freek{A} \otimes \freek{C}$ and $z : \freek{B} \otimes \freek{C}$.
If all three inputs have size $n$ it can be shown (e.g. using fractional edge covers~\cite{grohe:fractional}) that the output has size $O(n^{\frac{3}{2}})$ and indeed our simplification algorithm, sketched above, produces the output in $O(n^{\frac{3}{2}})$ worst-case time.
Traditional approaches using query planning cannot get below $O(n^2)$ worst-case time. 

More generally, for any given query and given sizes of input data we can consider the worst-case output size.
Roughly speaking, a \emph{worst-case optimal} algorithm solves the join problem in time proportional to the largest possible output size for an input of a given size.
This is realistically the best we can hope for since conjunctive queries are powerful enough to encode SAT-like problems.

Our simplification procedure, when applied to the special case of conjunctive queries, is indeed worst-case optimal. This is tricky to show since 
computation steps that compute a multiplicative subterm $v v'$ that \emph{eventually} yields $0$ are wasted: they contribute nothing to the output! This
is why it is important to choose the $v_i$ whose sum contains the fewest components above.

\section{Discussion}
\label{sec:discussion}

We have developed an algebraic theory based on mathematical modules and a number of associated categorical constructions as a simultaneously expressive, reasonable \emph{and} efficient purely functional programming framework for generalised relational query languages.  Its efficiency is based on
employing 
symbolic operators 
to retain data 
in \emph{un}normalised form at run time, exploiting their universal properties to perform optimising algebraic rewriting. 
Informally, the functions operate on quotient types modulo algebraic equivalences \emph{behaviourally}, 
but their efficient \emph{execution} 
crucially depends on 
the specifics of these representations.  

We have shown how controlled algebraic rewriting combined with efficient 
implementation of linear isomorphisms 
can be used to implement generalised database joins that perform asymptotically as well as can be hoped for. 
This hinges crucially on an extra element adjoined to free modules and finite maps, which has an intuitive interpretation both algebraically (as multiplicative unit for intersection) and as a wildcard (``compatible with anything'') and thus facilitates compactly representing infinite relations in combination with subtraction and tensor product.  

\subsection{Implementation}
The concepts we have presented are implemented as a Haskell library.
Modules are represented by a data type \verb|Space|.
Using the Data Kinds extension this will serve as a parameter for the type of elements:
\begin{verbatim}
data Vec r v = Zero | Add (Vec r v) (Vec r v)
             | Mul r (Vec r v) | Gen (Gen r v)
data family Gen r (v :: Space) :: Type
\end{verbatim}
An element of \verb|Vec r v| represents a term of the \verb|r|-module \verb|v|.
Such an element is built from linear combinations of generators, represented by the type family \verb|Gen r v|.
Most instances are simple, like tensor products:
\begin{verbatim}
data instance Gen r (u :*: v) = Tensor (Vec r u) (Vec r v)
\end{verbatim}
Finite maps, however, do not admit a both generic and efficient definition.
Instead, they are defined 
for each index type based on natural isomorphisms.
For example, the isomorphism $\fmapprodiso : (A \times B) \fmap U \iso A \fmap B \fmap U$ motivates the definition:
\begin{verbatim}
newtype instance Gen r ((a, b) :=> v) = CopowProd (Vec r (a :=> b :=> v))
\end{verbatim}
Thus, \verb|CopowProd| directly represents $\fmapprodiso^{-1}$, but since we are using \verb|newtype| it has no computational cost.

\paragraph{Related work}


Our modules and associative algebras provide a general, established and mathematically deep and well-studied reference framework for structures proposed in the literature, such as provenance semirings~\cite{semirings} and semiring dictionaries~\cite{shaikhha2021functional}.  For example, in provenance semirings
a $K$-relation is an element of $\free{K}{T_{a_1} \times \ldots \times T_{a_n}}$. Replacing $K$ by $R = \Nat[X]$, the free commutative semiring generated by $X$ yields queries evaluated over $R$ instead of $K$. 
This incorporates provenance for aggregations~\cite{amsterdamer2011provenance} since $\freek{T}$ provides folding over arbitrary modules.
Our extension with wildcards extends provenance from finite $K$-relations to infinite $K$-relations. Further, our approach accounts for probabilistic databases~\cite{probdb2007} as algebras $\free{\Real}{T}$, with real numbers serving as quasi-probabilities, as well as sets with negative multiplicity (see, e.g.,~\cite{loeb:negsets,caretteetal:symbdom}).



Linear algebra has previously been proposed as a framework for interpreting and implementing querying on databases.  Notably, typed matrix algebra~\cite{oliveira2017data,oliveira2012towards,oliveira2012typed,macedo2012typing} has been used to provide operations on matrix-shaped data (see also array programming languages such as Futhark~\cite{futhark} and Single Assignment C~\cite{Scholz:SaC}).  Note that our approach works on infinite-dimensional spaces, not only finite-dimensional ones, and it is essential to represent linear maps as functions rather than matrices. 
More closely related to our work, there are expressive linear algebra-based domain-specific languages/frameworks~\cite{kiselyov2018reconciling,shaikhha2019finally} that provide expressive database and data analytics operations, with evaluation to standard normalised data structures using Kiselyov's tagless final approach.   These do not support efficient data structures and execution of \emph{bi}linear maps, 
however, where we employ symbolic tensor products and efficient intersection, aided by compact maps for both expressiveness and 
efficiency.

The problem of efficient execution of relational database queries has been studied extensively, 
with joins
in particular 
posing a challenge. Traditional methods rely on decomposing $k$-ary joins into binary joins, 
carefully 
creating a good \emph{query plan}
in order 
to minimise the sizes of intermediate results 
and outputs~\cite{largedatabases}, while our approach instead hinges on sensible and efficient \emph{expression simplification} (see~\cite{carette:expsimp} for a formal treatment of this problem).
While it is possible to evaluate \emph{acyclic} join queries in time linear in the size of the query, the input, and the output~\cite{willard2002,join-acyclic} and there are methods that deal with ``almost'' acyclic join queries~\cite{evaltreedecomp}, this is 
infeasible for cyclic queries such as the triangle query, as deciding whether the output is empty is NP-hard~\cite{maier1981complexity}.
To quantify 
how good a query algorithm is, the notion of \emph{worst-case optimal complexity} was developed~\cite{ngo2012worst,ngo2018worst}, focussing on the \emph{data complexity} of the problem, which ignores the size of the query itself~\cite{complexity}. A join algorithm is 
\emph{worst-case optimal} if it executes in time linear in $N$ and $O$, where $O$ is the maximal size
of an output of the join query applied to input relations whose sizes sum to $N$. 

We have drawn on a great number of categorical concepts, not only to drive generalisation and identify suitable compositional constructs 
that make for a pleasant framework with strong and useful properties, but also to exploit term constructors as symbolic data structures and exception-less algebraic equalities for their efficient 
execution. 
It is difficult to do justice to the many works in category theory, abstract algebra and categorical functional programming we have built on. 
Most closely related is~\cite{gibbons2018relational}, which presents a framework for relational algebra based on the adjunctions that generate it.

\paragraph{Future work}
While logic programming does well in handling positive queries, negation is typically handled in an ad-hoc and unsatisfactory way, for example through \emph{negation as failure}. In our setting, data with \emph{negative} multiplicity is handled no different from data in \emph{positive} quantity: can this give a more satisfactory treatment of negation in logic programming?
Further, in the current form, our setting is quite conservative in that it is not capable of describing recursive queries, even if they are ultimately well-behaved. A possible solution to this shortcoming would be to study modules which are somehow topological (e.g., appropriately ordered, or equipped with a norm or inner product), as this could give us access to fixed point theorems directly associated with the semantics of recursion.

\paragraph{Acknowledgements.}
\begin{small}
This work was made possible by Independent Research Fund Denmark grant \emph{FUTHARK: Functional Technology for High-performance Architectures} and DFF--International Postdoc 0131-00025B.  We greatly appreciate and thank the three anonymous referees for their comments and recommendations. 
\end{small}

\bibliographystyle{eptcs}
\bibliography{p}

\appendix
\section{Appendix}

We provide definitions of rings and modules.  See the literature (for example~\cite{wisbauer1991foundations}) for more information.

\subsection{Rings}

\begin{definition}[Ring]
\label{ring}
A \emph{commutative ring} $\K$ is an algebraic structure \\ $\K = (\car \K, +, -, 0, \cdot, 1)$ such that for all $x, y, z \in \car \K$:
\begin{eqnarray}
x + (y + z) & = & (x + y) + z \\
x + y & = & y + x \\
x + 0 & = & x \\
x + (-x) = 0 \\
0 \cdot x & = & 0 \\
1 \cdot x & = & x \\
x \cdot (y \cdot z) & = & (x \cdot y) \cdot z \\
x \cdot (y + z) & = & (x \cdot y) + (x \cdot z) \\
x \cdot y & = & y \cdot x
\end{eqnarray}
\end{definition}

Typical examples of rings are $\Int$, $\Rat$, $\Real$ and $\Complex$ with the usual addition and multiplication operations.

\subsection{Modules}
\begin{definition}[Module]
\label{modoule}
A \emph{module} $V$ over commutative ring $\K$ is an algebraic structure $V = (\car V, +, 0, \cdot)$ such that for all $r, s \in \car \K$, $u, v, w \in \car V$:
\begin{eqnarray}
u + (v + w) & = & (u + v) + w \\
u + v & = & v + u \\
u + 0 & = & u \\
0 \cdot u & = & 0 \\
1 \cdot u & = & u \\
r \cdot (s \cdot u) & = & (r \cdot s) \cdot u \\
r \cdot (u + v) & = & (r \cdot u) + (r \cdot v) \\
(r + s) \cdot u & = & (r \cdot u) + (s \cdot u)
\end{eqnarray}
\end{definition}

All modules support negation as well, defined by $-u = (-1) \cdot u$.
If $\K$ is a field then $V$ is a vector space in the usual sense.

Given two modules $U$ and $V$, a \emph{linear map} between them is a function $f : U \to V$ of the underlying sets such that
\begin{eqnarray*}
f(u + v) &=& f(u) + f(v) \\
f(r \cdot u) &=& r \cdot f(u)
\end{eqnarray*}

Given three modules $U$, $V$ and $W$, a \emph{bilinear map} between them is a function $f : U \times V \to W$ of the underlying sets such that
\begin{eqnarray*}
f(u_1 + u_2, v) &=& f(u_1, v) + f(u_2, v) \\
f(u, v_1 + v_2) &=& f(u, v_1) + f(u, v_2) \\
f(r \cdot u, v) &=& r \cdot f(u, v) \\
f(u, r \cdot v) &=& r \cdot f(u, v)
\end{eqnarray*}
In other words a bilinear map is linear in each of its arguments.

\subsection{Biproducts}
\label{appendix:biproduct}
The biproduct satisfies the universal property of binary products: there are linear maps $p_1 : U \oplus V \to U$ and $p_2 : U \oplus V \to V$ such that for any module $W$ with linear maps $p_1' : W \to U$ and $p_2' : W \to V$ there is a unique linear map $(p_1', p_2') : W \to U \oplus V$ such that $p_1 \comp (p_1', p_2') = p_1'$ and $p_2 \comp (p_1', p_2') = p_2'$.
The $p$'s are projections and $(p_1', p_2')$ is the pair constructor.
Operationally:
\begin{eqnarray*}
p_1(u, v) &=& u \\
p_2(u, v) &=& v \\
(p_1', p_2')(w) &=& (p_1'(w), p_2'(w))
\end{eqnarray*}

This universal property gives a way to define linear maps \emph{into} the biproduct by giving the result of projecting each component.
In other words, definition by copattern matching.
For instance
\begin{eqnarray*}
f &:& V \to V \oplus V \\
p_1(f(v)) &=& v \\
p_2(f(v)) &=& 3 v
\end{eqnarray*}

The biproduct also satisfies the universal property of binary coproducts, obtained by simply reversing the direction of all arrows: there are linear maps $i_1 : U \to U \oplus V$ and $i_2 : V \to U \oplus V$ such that for any module $W$ with linear maps $i_1' : U \to W$ and $i_2' : V \to W$ there is a unique linear map $[i_1', i_2'] : U \oplus V \to W$ such that $[i_1', i_2'] \comp i_1 = i_1'$ and $[i_1', i_2'] \comp i_2 = i_2'$.

The $i$'s are injections and $[i_1', i_2']$ is case analysis.
Operationally:
\begin{eqnarray*}
i_1(u) &=& (u, 0) \\
i_2(v) &=& (0, v) \\
{[}i_1', i_2'{]}(u, v) &=& i_1'(u) + i_2'(v)
\end{eqnarray*}

This universal property gives a way to define linear maps \emph{out of} the biproduct by explaining what happens to either injection.
In other words, definition by pattern matching.
For instance
\begin{eqnarray*}
f &:& V \oplus V \to V \\
f(i_1(v)) &=& 2 v \\
f(i_2(v)) &=& v
\end{eqnarray*}

From a programming point of view biproducts are an unfamiliar construction: a data type that supports both pattern matching and copattern matching.
They can be treated as either a sum or a product type depending on which is most convenient.

Finally, we describe the canonical algebra structure on the biproduct.
Just like for free modules, the motivation for the definition is intersection of multisets.
Suppose $U$ and $V$ are algebras.
Then so is $U \oplus V$.
The definition of $1_{U \oplus V}$ uses copattern matching while $\cdot$ and $\weight$ use pattern matching.
\begin{eqnarray*}
p_1(1_{U \oplus V}) &=& 1_U \\
p_2(1_{U \oplus V}) &=& 1_V \\
i_1(u) \cdot i_1(u') &=& i_1(u \cdot u') \\
i_1(u) \cdot i_2(v') &=& 0 \\
i_2(v) \cdot i_1(u') &=& 0 \\
i_2(v) \cdot i_2(v') &=& i_2(v \cdot v') \\
\weight i_1(u) &=& \weight u \\
\weight i_2(v) &=& \weight v
\end{eqnarray*}

An element $(u, v) \in U \oplus V$ can be thought of as a record value $\{ 1 \mapsto u, 2 \mapsto v \}$.  Biproducts generalise straightforwardly from tuples to \emph{records} via $\oplus_{i \in I} U_i$, where $I$ is an index set $I$ of field names $i$, whose associated values come from a module $U_i$ that depends on $i$.

\subsection{Naive intersection}
\label{appendix:naiveintersection}
The join of two elements of the same module is their intersection, which is given by multiplication.
\small
\[
x' \cdot y' = (\inj{a} \otimes \inj{1} \otimes 1 + \inj{b} \otimes \inj{2} \otimes 1 + \inj{c} \otimes \inj{3} \otimes 1) \cdot (1 \otimes \inj{2} \otimes \inj{p} + 1 \otimes \inj{3} \otimes \inj{q} + 1 \otimes \inj{4} \otimes \inj{r}) \\
\]
\normalsize
Using distributivity we expand the multiplication.
\scriptsize
\begin{eqnarray*}
& &(\inj{a} \otimes \inj{1} \otimes 1) \cdot (1 \otimes \inj{2} \otimes \inj{p}) + (\inj{a} \otimes \inj{1} \otimes 1) \cdot (1 \otimes \inj{3} \otimes \inj{q}) + (\inj{a} \otimes \inj{1} \otimes 1) \cdot (1 \otimes \inj{4} \otimes \inj{r}) \\
&+&(\inj{b} \otimes \inj{2} \otimes 1) \cdot (1 \otimes \inj{2} \otimes \inj{p}) + (\inj{b} \otimes \inj{2} \otimes 1) \cdot (1 \otimes \inj{3} \otimes \inj{q}) + (\inj{b} \otimes \inj{2} \otimes 1) \cdot (1 \otimes \inj{4} \otimes \inj{r}) \\
&+&(\inj{c} \otimes \inj{3} \otimes 1) \cdot (1 \otimes \inj{2} \otimes \inj{p}) + (\inj{c} \otimes \inj{3} \otimes 1) \cdot (1 \otimes \inj{3} \otimes \inj{q}) + (\inj{c} \otimes \inj{3} \otimes 1) \cdot (1 \otimes \inj{4} \otimes \inj{r})
\end{eqnarray*}
\normalsize
Next we exploit that $\cdot$ works component-wise on tensor products.
\scriptsize
\begin{eqnarray*}
& &(\inj{a} \cdot 1) \otimes (\inj{1} \cdot \inj{2}) \otimes (1 \cdot \inj{p}) + (\inj{a} \cdot 1) \otimes (\inj{1} \cdot \inj{3}) \otimes (1 \cdot \inj{q}) + (\inj{a} \cdot 1) \otimes (\inj{1} \cdot \inj{4}) \otimes (1 \cdot \inj{r}) \\
&+&(\inj{b} \cdot 1) \otimes (\inj{2} \cdot \inj{2}) \otimes (1 \cdot \inj{p}) + (\inj{b} \cdot 1) \otimes (\inj{2} \cdot \inj{3}) \otimes (1 \cdot \inj{q}) + (\inj{b} \cdot 1) \otimes (\inj{2} \cdot \inj{4}) \otimes (1 \cdot \inj{r}) \\
&+&(\inj{c} \cdot 1) \otimes (\inj{3} \cdot \inj{2}) \otimes (1 \cdot \inj{p}) + (\inj{c} \cdot 1) \otimes (\inj{3} \cdot \inj{3}) \otimes (1 \cdot \inj{q}) + (\inj{c} \cdot 1) \otimes (\inj{3} \cdot \inj{4}) \otimes (1 \cdot \inj{r})
\end{eqnarray*}
\normalsize
The multiplications now all involve only generators and $1$, so they can all be simplified.
\begin{eqnarray*}
& &\inj{a} \otimes 0 \otimes \inj{p} + \inj{a} \otimes 0 \otimes \inj{q} + \inj{a} \otimes 0 \otimes \inj{r} \\
&+&\inj{b} \otimes \inj{2} \otimes \inj{p} + \inj{b} \otimes 0 \otimes \inj{q} + \inj{b} \otimes 0 \otimes \inj{r} \\
&+&\inj{c} \otimes 0 \otimes \inj{p} + \inj{c} \otimes \inj{3} \otimes \inj{q} + \inj{c} \otimes 0 \otimes \inj{r}
\end{eqnarray*}
All tensor products with a $0$ component can be eliminated due to linearity.
This leaves:
\[
\inj{b} \otimes \inj{2} \otimes \inj{p} + \inj{c} \otimes \inj{3} \otimes \inj{q}
\]

\section{Simplification example} 
\label{appendix:simplification-example}
For primitive types we will simply write the simplified form as $y = \sum_i (a_i \mapsto u_i)$ with the understanding that each $a : A$ occurs at most once, and computing $y(a)$ can be done using work linear in the size of $a$ (which for finite precision integers would be constant). 
To see all this in action consider the term
\begin{eqnarray*}
x &:& (\Str \times (\Str + \Nat)) \fmap \K \\
x &=& ((a, \inl(p)) \mapsto 1) + ((b, \inr(4)) \mapsto 1) + ((a, \inr(3)) \mapsto 1) + ((a, \inl(p)) \mapsto 1)
\end{eqnarray*}
The simplification proceeds as follows:
\small
\begin{eqnarray*}
x &=& ((a, \inl(p)) \mapsto 1) + ((b, \inr(4)) \mapsto 1) + ((a, \inr(3)) \mapsto 1) + ((a, \inl(p)) \mapsto 1) \\
& & \qquad \text{apply $\fmapprodiso$} \\
&=& \fmapprodiso^{-1}((a \mapsto \inl(p) \mapsto 1) + (b \mapsto \inr(4) \mapsto 1) + (a \mapsto \inr(3) \mapsto 1) + (a \mapsto \inl(p) \mapsto 1)) \\
& & \qquad \text{simplify outer finite map using method for strings} \\
&=& \fmapprodiso^{-1}((a \mapsto (\inl(p) \mapsto 1) + (\inr(3) \mapsto 1) + (\inl(p) \mapsto 1)) + (b \mapsto \inr(4) \mapsto 1)) \\
& & \qquad \text{apply $\fmapsumiso$} \\
&=& \fmapprodiso^{-1}((a \mapsto \fmapsumiso^{-1}((p \mapsto 1, 0) + (0, 3 \mapsto 1) + (p \mapsto 1, 0))) + (b \mapsto \fmapsumiso^{-1}(0, 4 \mapsto 1))) \\
& & \qquad \text{simplify biproduct by adding pairwise} \\
&=& \fmapprodiso^{-1}((a \mapsto \fmapsumiso^{-1}((p \mapsto 1) + (p \mapsto 1), 3 \mapsto 1)) + (b \mapsto \fmapsumiso^{-1}(0, 4 \mapsto 1))) \\
& & \qquad \text{simplify inner finite map using methods for strings and integers} \\
&=& \fmapprodiso^{-1}((a \mapsto \fmapsumiso^{-1}(p \mapsto 2, 3 \mapsto 1)) + (b \mapsto \fmapsumiso^{-1}(0, 4 \mapsto 1)))
\end{eqnarray*}
\normalsize
Using this method the simplified form of a finite map can be found in linear time.
Compact maps can be dealt with just as easily by exploiting 
that $A \cmap U \iso (A \fmap U) \oplus U$. 
These isomorphisms are the basis of generic tries~\cite{hehi2013,tries}.
%

\end{document}